\documentclass[letterpaper]{article}

\usepackage[T1]{fontenc}

\usepackage{geometry}
\usepackage{setspace}
\usepackage{achemso}

\usepackage{graphicx}
\usepackage{float}
\newfloat{scheme}{htbp}{los}
\floatname{scheme}{Scheme}
\floatname{chart}{Chart}
\newfloat{graph}{htbp}{loh}

\usepackage{chemformula} 
\usepackage[version = 4]{mhchem} 
\usepackage{amsmath}
\usepackage{bm}
\usepackage[colorlinks,linkcolor=blue,anchorcolor=blue,citecolor=blue]{hyperref}

\setkeys{acs}{articletitle=true}

\usepackage{authblk}
\author[1]{Xinyuan Jiang}
\author[1,2]{Jian Wu}
\author[1,3]{Weiyi Pan*}
\affil[1]{State Key Laboratory of Low Dimensional Quantum Physics, Department of Physics, Tsinghua University, Beijing 100084, China}
\affil[2]{Frontier Science Center for Quantum Information, Beijing, China}
\affil[3]{Institute for Theoretical Physics, University of Regensburg, 93040 Regensburg, Germany}

\title{Switching between Antiferromagnetic and Ferromagnetic Skyrmions in Two-Dimensional Magnets}
\date{*Email: weiyi.pan@physik.uni-regensburg.de}

\begin{document}

\maketitle

\begin{abstract}
Antiferromagnetic (AFM) and ferromagnetic (FM) skyrmions possess unique advantages for spintronic applications. AFM skyrmions eliminate the skyrmion Hall effect and exhibit fast dynamics, whereas FM skyrmions are easier to nucleate and manipulate. However, realizing a transition between AFM and FM skyrmions within the same two-dimensional (2D) material has remained elusive. Here, using first-principles calculations and atomistic spin simulations on the Janus monolayer $\ce{Cr2Ge2Te3S3}$, we demonstrate that strain-driven modulation of magnetic interactions enables switching between AFM and FM skyrmion phases. A compressive strain of $-3\%$ induces an AFM ground state hosting AFM skyrmions, while a tensile strain of $+2\%$ drives the system into a FM skyrmion phase.
Moreover, under an out-of-plane magnetic field, FM skyrmions are rapidly transformed into a uniform FM phase, while AFM skyrmions transform into AFM bimerons under stronger fields. These findings establish a framework for controllable transitions between topological magnetic states in a single 2D material.
\end{abstract}

\section*{Keywords}

antiferromagnetic skyrmions, ferromagnetic skyrmions, strain, Janus monolayer, first-principles

\begin{figure}
\centering
\includegraphics[width=0.3\textwidth]{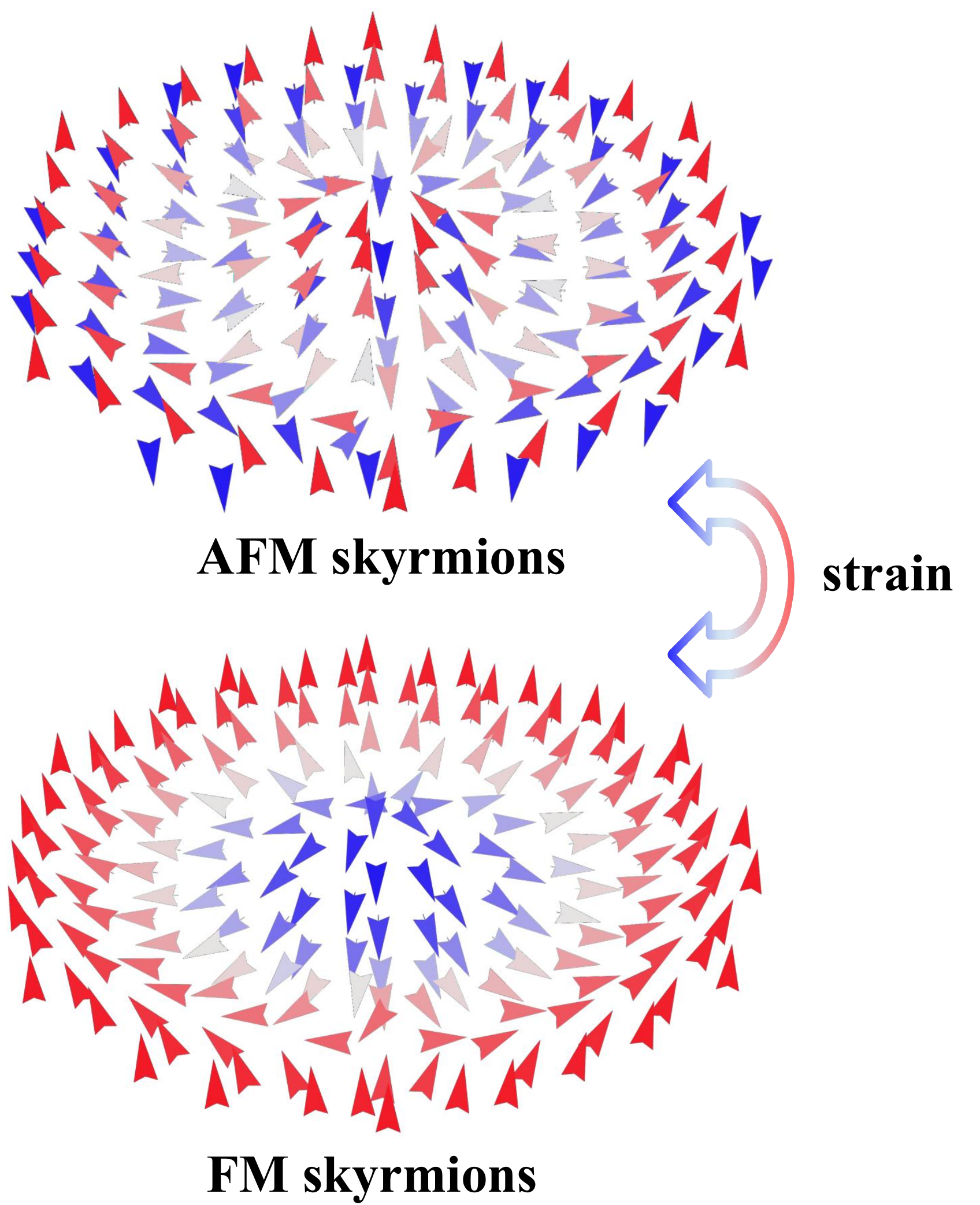}
\end{figure}

\section{Introduction}

Topological magnetic textures, such as magnetic skyrmions, have emerged as a central topic in condensed matter physics owing to their nontrivial topology, nanoscale size, and potential applications in next-generation spintronic devices \cite{doi:10.1126/science.1166767, Yu2010, Fert2017}. Depending on the magnetic background in which they exist, skyrmions can be classified into ferromagnetic (FM) and antiferromagnetic (AFM) types. FM skyrmions have been extensively studied and experimentally realized in a variety of materials, benefiting from their finite magnetization that enables direct detection and efficient manipulation by magnetic fields or spin-polarized currents \cite{Zhang_2023, Zhang2018, Wang2022}. However, these same characteristics also give rise to intrinsic limitations: the nonzero net magnetization and topological charge induce a pronounced skyrmion Hall effect (SkHE), leading to transverse motion under current-driven dynamics and complicating precise trajectory control \cite{Nagaosa2013, Jiang2017, Litzius2017}. In addition, dipolar interactions tend to increase the skyrmion size, posing challenges for device miniaturization \cite{Büttner2018}. In contrast, AFM skyrmions, composed of two antiparallel FM sublattices coupled antiferromagnetically, exhibit complementary properties. Their compensated magnetization and topological charge suppress the SkHE, resulting in strictly linear motion and enhanced dynamical stability \cite{GOBEL20211, PhysRevLett.116.147203, Zhang2016, Gomonay2018, Jungwirth2016, Šmejkal2018}. This intrinsic complementarity between FM and AFM skyrmions highlights the importance of considering both types of topological textures on equal footing. In recent years, the rapid progress in two-dimensional (2D) magnetic materials has further stimulated intensive research on both FM and AFM skyrmion crystals, driven by their promise for highly integrated and energy-efficient topological spintronic devices \cite{PhysRevB.109.214405, PhysRevB.111.054423, PhysRevB.110.144408, 753s-8nx4}.

Given the complementary advantages of different topological magnetic states, an important objective for functional spintronic applications is to achieve controlled regulation and interconversion between distinct topological textures. A variety of external stimuli, including magnetic fields \cite{PhysRevB.101.094420, PhysRevB.101.060404, doi:10.1021/acs.nanolett.2c00836}, electric fields \cite{Sun2020, D3MH00572K, PhysRevLett.125.037203}, and thermal excitation \cite{PhysRevB.104.054420, Amoroso2020, doi:10.1021/acs.nanolett.3c05024}, have been demonstrated to effectively manipulate skyrmions and related textures. However, in most reported cases, such manipulations are performed within a fixed magnetic background, for example, transforming one FM topological texture into another, such as from FM skyrmions to FM bimerons \cite{doi:10.1021/acs.nanolett.5c05775, 753s-8nx4}, without altering the underlying FM exchange framework. In contrast, a direct switching between FM and AFM topological states remains largely unexplored. The difficulty is that such a transition requires a change in the sign of the dominant exchange interaction, which is tightly bound to the electronic structure and crystal geometry of the material and therefore not easily tunable. Nevertheless, considering the complementary transport and dynamical properties of FM and AFM skyrmions, the ability to switch between FM and AFM skyrmion phases would represent a critical step toward multifunctional and reconfigurable topological devices, enabling different operational modes within a single material platform.

In this work, taking the Janus monolayer $\ce{Cr2Ge2Te3S3}$ as a representative model system, we demonstrate that a switching between FM and AFM skyrmion phases can be realized through strain-induced modulation of magnetic interactions. Based on first-principles calculations and atomistic spin simulations, we show that the application of strain tunes the underlying magnetic background from FM to AFM coupling, thereby enabling the emergence of distinct skyrmion crystals with different topological and dynamical characteristics within the same material platform. Our analysis reveals that these textures originate from the competition among the Heisenberg exchange, Dzyaloshinskii-Moriya interaction (DMI), and magnetic anisotropy, whose relative balance is continuously modulated by strain. Under external magnetic fields, the FM skyrmion phase can be easily reshaped or annihilated, whereas the AFM skyrmion state exhibits robustness owing to its compensated spin structure. These findings open a previously unexplored route toward strain-mediated interconversion between FM and AFM topological magnetic states, offering a promising pathway for the development of multifunctional and reconfigurable topological spintronic devices.

\section{Results and discussion}

\begin{figure}
\centering
\includegraphics[height=4cm]{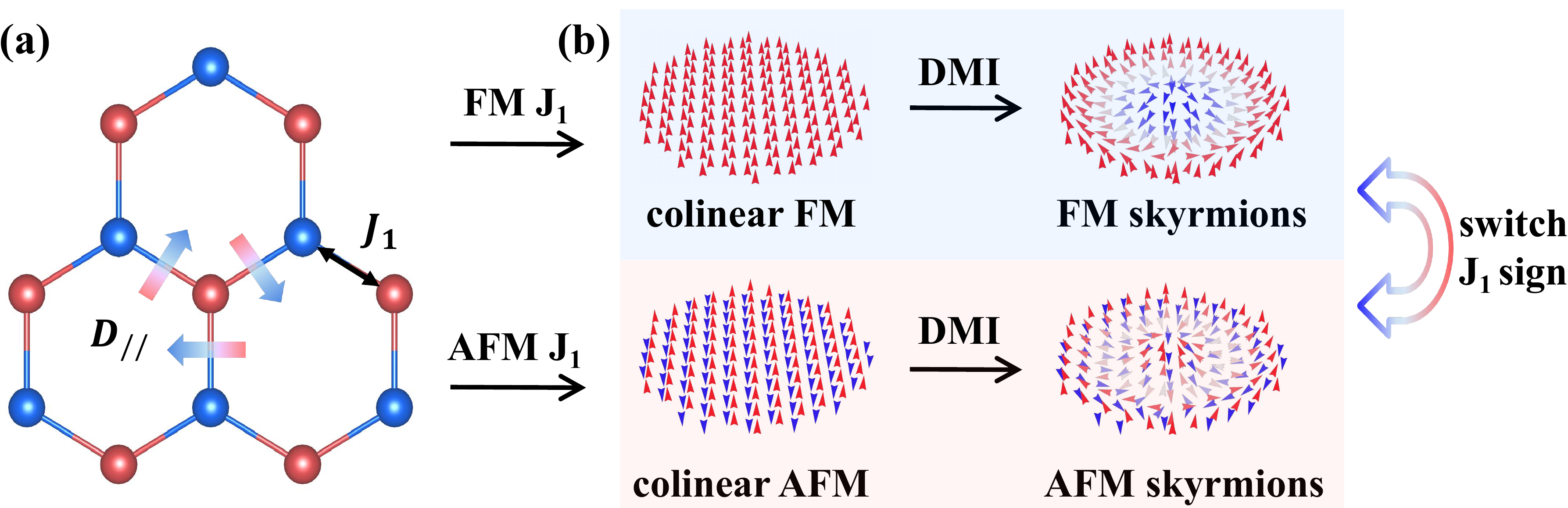}
\caption{\label{fig:figure1} 
General mechanism for the transition between FM and AFM skyrmions.
(a) Schematic illustration of a 2D hexagonal lattice composed of two magnetic sublattices (blue and red atoms). The nearest-neighbor exchange interaction $J_1$ connects spins on adjacent sites, while the in-plane DMI $D_{//}$ is indicated by arrows (anticlockwise chirality shown as an example).
(b) Schematic transition of magnetic textures. Under finite $D_{//}$, a collinear FM state can transform into a FM skyrmion state, while a collinear AFM state can transform into an AFM skyrmion state. The transition between FM and AFM skyrmions can be achieved by switching the sign of $J_1$.
}
\end{figure}

From a physical perspective, the formation of skyrmions generally originates from the competition between the Heisenberg exchange interaction and the DMI \cite{PhysRev.120.91, DZYALOSHINSKY1958241, PhysRevLett.108.017206}. While the exchange interaction favors collinear spin alignment, the DMI promotes chiral noncollinear spin textures, and their interplay stabilizes topological spin configurations in magnetic systems. To illustrate this principle, we consider a prototypical 2D hexagonal lattice, as shown in Fig.~\ref{fig:figure1}. When the nearest-neighbor exchange interaction ($J_1$) is FM, the presence of finite DMI can twist the spins into a chiral configuration and stabilize a FM skyrmion state. In contrast, when $J_1$ is AFM, the magnetic lattice naturally divides into two antiparallel sublattices, and the resulting chiral spin texture corresponds to an AFM skyrmion.
This suggests a general route for realizing the transition between FM and AFM skyrmions. As illustrated in Fig.~\ref{fig:figure1}, once a finite DMI exists to stabilize chiral spin textures, tuning $J_1$ from FM to AFM can transform the magnetic background and consequently drive the evolution between these two types of skyrmions.

\subsection{Strain-tunable magnetic ground states}

\begin{figure}
\centering
\includegraphics[height=10cm]{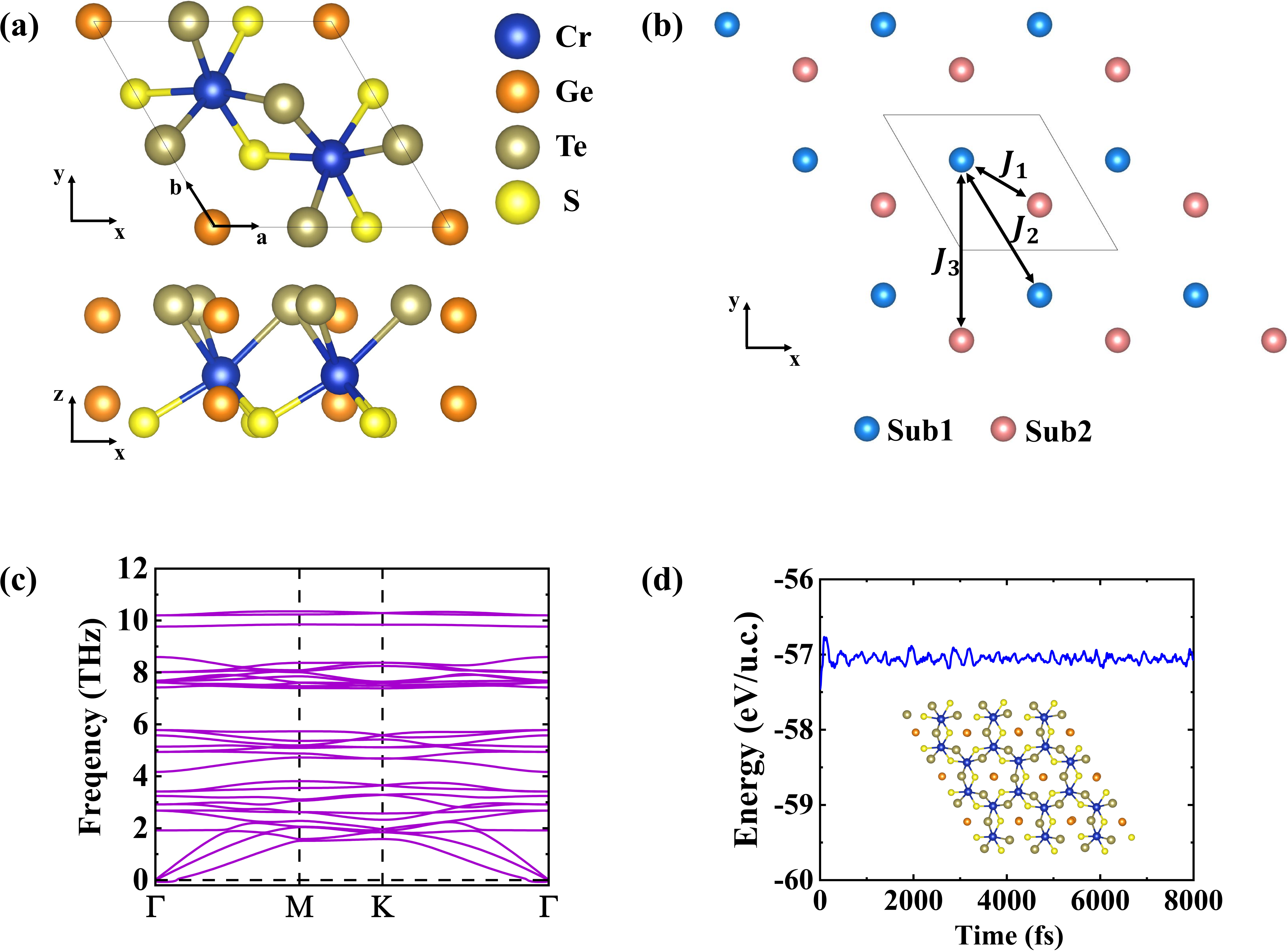}
\caption{\label{fig:figure2} 
(a) Crystal structures of $\ce{Cr2Ge2Te3S3}$. The upper panel shows the top view along the $z$-axis, and the lower panel presents the side view perpendicular to the $z$-axis.
(b) Schematic representation of Heisenberg exchange interactions, with first-nearest neighbor $J_1$, second-nearest neighbor $J_2$, and third-nearest neighbor $J_3$.
(c) Phonon spectra of $\ce{Cr2Ge2Te3S3}$.
(d) AIMD simulations of monolayer $\ce{Cr2Ge2Te3S3}$ at 300 K. The energy is given, where the inset shows the crystal structure after 8000 fs.
}
\end{figure}

\begin{figure}
\includegraphics[width=0.9\textwidth]{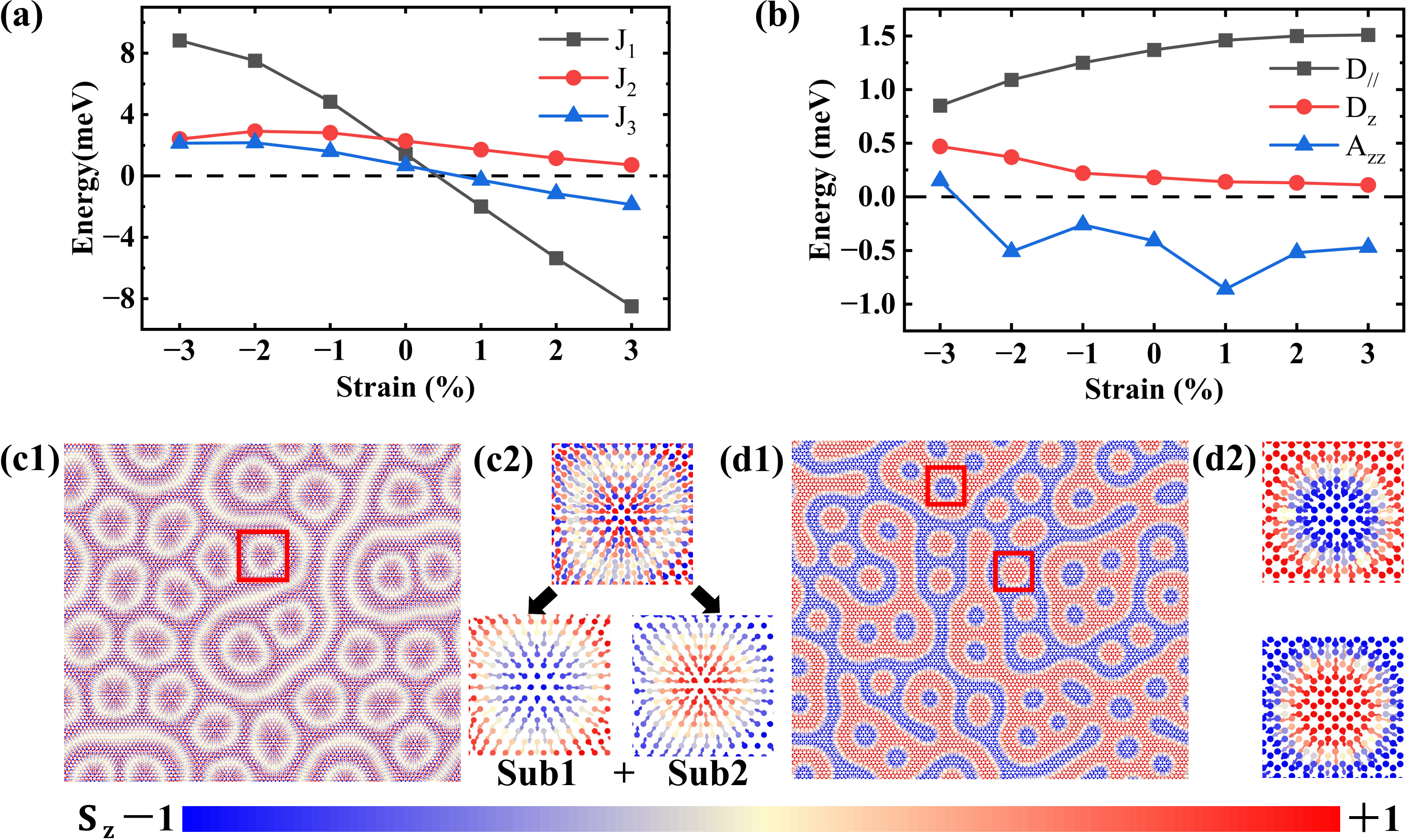}
\caption{\label{fig:figure3} 
Strain dependence of the magnetic parameters, magnetic phase diagrams, and detailed spin configurations of monolayer $\ce{Cr2Ge2Te3S3}$.
(a) Variations of the Heisenberg exchange ($J_1$, $J_2$, and $J_3$) as functions of biaxial strain.
(b) Variations of the in-plane ($D_{//}$) and out-of-plane ($D_{z}$) components of the DMI, together with the SIA ($A_{zz}$), under the same strain conditions. For simplicity, all spin vectors are normalized to $|\mathbf{S}|=1$ in our calculations.
(c1) Enlarged view of the magnetic texture at $-3\%$ strain, showing the formation of AFM skyrmions.
(c2) Magnified region highlighted by the red box in (c1), where the AFM skyrmion can be decomposed into two interpenetrating sublattices, Sub1 and Sub2, corresponding to FM skyrmions with opposite spin orientations (spin-up and spin-down, respectively).
(d1) Enlarged view of the magnetic texture at $+2\%$ strain, exhibiting FM skyrmions.
(d2) Further magnification of the red-boxed regions in (d1), illustrating two types of FM skyrmions with opposite spin orientations.
}
\end{figure}

After clarifying the conditions for the transition from FM to AFM skyrmions, we take $\ce{Cr2Ge2Te3S3}$ as a representative 2D magnetic material to demonstrate how this concept can be realized in a realistic system.
The crystal structure of monolayer $\ce{Cr2Ge2Te3S3}$ is illustrated in Fig.~\ref{fig:figure2}(a). The system crystallizes in the polar Janus structure with the space group $P3m1$ (No. 157). It can be viewed as a derivative of $\ce{CrGeTe3}$, in which the bottom Te layer is replaced by S atoms. This asymmetric configuration breaks inversion symmetry, giving rise to a DMI between neighboring Cr atoms. The optimized in-plane lattice constant, defined as the distance between the two Ge atoms in the same layer, is 6.35 \AA. The Cr atoms form a hexagonal lattice with two inequivalent Cr sites per unit cell, as shown in Fig.~\ref{fig:figure2}(b). Each Cr atom is octahedrally coordinated by three Te and three S atoms, forming a distorted octahedron. The mixed chalcogen coordination produces a crystal-field splitting of the Cr $3d$ orbitals, where the octahedral ligand field separates the states into fully occupied threefold-degenerate $t_{2g}$ levels and empty twofold-degenerate $e_g$ levels. We evaluate the formation energy ($E_f$) of monolayer $\ce{Cr2Ge2Te3S3}$, which is defined as 
\begin{eqnarray}
\begin{aligned}
    E_f=\frac{E_{\ce{Cr2Ge2Te3S3}}-2E_{\ce{Cr}}-2E_{\ce{Ge}}-3E_{\ce{Te}}-3E_{\ce{S}}}{10}
    \label{eq:E_f}
\end{aligned}
\end{eqnarray}
where $E_{\ce{Cr2Ge2Te3S3}}$ is the total energy of monolayer $\ce{Cr2Ge2Te3S3}$ per unit cell, and $E_{\ce{X}}$ (X = Cr, Ge, Te, S) are the energies per atom for corresponding elemental bulk phases. Specifically, Cr crystallizes in a body-centered cubic Phase, Ge in a diamond phase, Te in a trigonal structure, and S in an orthorhombic phase. The calculated $E_f$ of $-0.17$ eV per atom indicates that $\ce{Cr2Ge2Te3S3}$ is energetically and chemically stable. The phonon spectrum is shown in Fig.~\ref{fig:figure2}(c), obtained using the PHONOPY package, exhibits no imaginary frequencies, confirming the dynamical stability of the structure. Furthermore, AIMD simulations performed at 300 K in Fig.~\ref{fig:figure2}(d) show only small total-energy fluctuations and well-preserved atomic configurations throughout the simulation, verifying its thermal robustness at room temperature.

To gain further insight into the magnetic interactions, we constructed a spin Hamiltonian for monolayer $\ce{Cr2Ge2Te3S3}$:
\begin{eqnarray}
\begin{aligned}
    H=&J_1\sum_{\langle i,j\rangle}{\mathbf{S}_i}\cdot \mathbf{S}_j+
    J_2\sum_{\langle\langle i,j\rangle\rangle}\mathbf{S}_i\cdot \mathbf{S}_j+
    J_3\sum_{\langle\langle\langle i,j\rangle\rangle\rangle}\mathbf{S}_i\cdot \mathbf{S}_j\\
    &+A_{zz}\sum_i (S_{i}^{z})^2+
    \sum_{\langle i,j \rangle}\mathbf{D}_{ij}\cdot (\mathbf{S}_i \times \mathbf{S}_j)
    \label{eq:Hamiltonian}
\end{aligned}
\end{eqnarray}
In this equation, $\mathbf{S}_{i(j)}$ denotes the normalized spin vector of the $i$-th ($j$-th) Cr atom. $S_{i}^{z}$ denotes the $z$-component of the $i$-th spin. $J_1$, $J_2$ and $J_3$ correspond to the Heisenberg exchange parameters of the first, second, and third nearest neighbors (1NN, 2NN, and 3NN), respectively, as shown in Fig.~\ref{fig:figure2}(b). The summations $\langle i,j\rangle$, $\langle\langle i,j\rangle\rangle$, and $\langle\langle\langle i,j\rangle\rangle\rangle$ run over the 1NN, 2NN, and 3NN spin pairs, respectively. $A_{zz}$ is the single-ion anisotropy (SIA) energy. $\mathbf{D}_{ij}$ is the DMI between the $i$-th and $j$-th Cr atoms. Because the structure possesses a mirror plane perpendicular to the 1NN Cr-Cr bond, the DMI vector can be expressed as $\mathbf{D}_{ij}=D_{//}(\bm{\hat{u}}_{ij}\times \bm{\hat{z}})+D_z \bm{\hat{z}}$, where $\bm{\hat{u}}_{ij}$ represents the unit vector pointing along the Cr-Cr bond and $\bm{\hat{z}}$ represents the unit vector along the $z$ direction.

Using the spin Hamiltonian described above, the magnetic parameters are obtained by mapping the total energies of representative spin configurations. For the unstrained monolayer, the exchange couplings $J_1$, $J_2$, and $J_3$ are all relatively small and AFM, indicating weak overall exchange interaction that are unfavorable for the stabilization of magnetic orders. In contrast, the DMI is remarkably strong due to the broken inversion symmetry of the lattice. The magnetic coupling strength can be further tuned by strain. Upon applying biaxial strain, the magnetic exchange parameters exhibit systematic and contrasting trends. As shown in Fig.~\ref{fig:figure3}(a), $J_1$ is the most strain-sensitive parameter: compressive strain strengthens its AFM character, whereas tensile strain reverses it to a FM interaction. From a microscopic perspective, the compress-induced FM-to-AFM transition in $J_1$ can be ascribed to the strengthening of direct AFM exchange between Cr atoms, resulting from the contraction of the Cr-Cr bond distance. $J_2$, $J_3$, and DMI also vary with strain, but their changes are relatively moderate compared with $J_1$. These results indicate that strain primarily tunes the balance of magnetic exchange interactions, which provides a natural mechanism for switching between distinct magnetic ground states.

To gain deeper insights into the strain-dependent evolution of spin textures, we carry out LLG simulations on large supercells. The results indicate that strain can effectively modify the magnetic configurations, enabling the emergence of both FM and AFM skyrmions. Under compressive strain of $-3\%$, the system develops a complex AFM labyrinth domain structure interspersed with multiple isolated AFM skyrmions, as shown in Fig.~\ref{fig:figure3}(c1). A magnified view in Fig.~\ref{fig:figure3}(c2) reveals that each AFM skyrmion consists of two FM skyrmions coupled antiferromagnetically on interpenetrating sublattices. Unlike synthetic AFM skyrmions \cite{Chen2020, Zhang2016, Legrand2020, Duine2018}, which require two FM layers coupled through interlayer AFM exchange, the AFM skyrmions in our system are intrinsically realized within a single atomic layer. When the tensile strain is increased to $+2\%$, the AFM spin-spiral texture evolves into a FM skyrmion lattice, as shown in Fig.~\ref{fig:figure3}(d1). This transformation arises from the delicate balance among the competing exchange interactions, DMI, and magnetic anisotropy, which collectively determine the characteristic skyrmion size and stability. Remarkably, these results demonstrate that both AFM and FM skyrmions can be engineered within the same 2D Janus system through strain alone, providing an efficient and tunable route toward strain-controlled skyrmionics. In the following sections, we focus on the representative cases of $-3\%$ and $+2\%$ strain to elucidate the microscopic origins of their distinct spin configurations.

\subsection{Microscopic origin of AFM and FM skyrmions}

\begin{figure}
\centering
\includegraphics[width=0.9\textwidth]{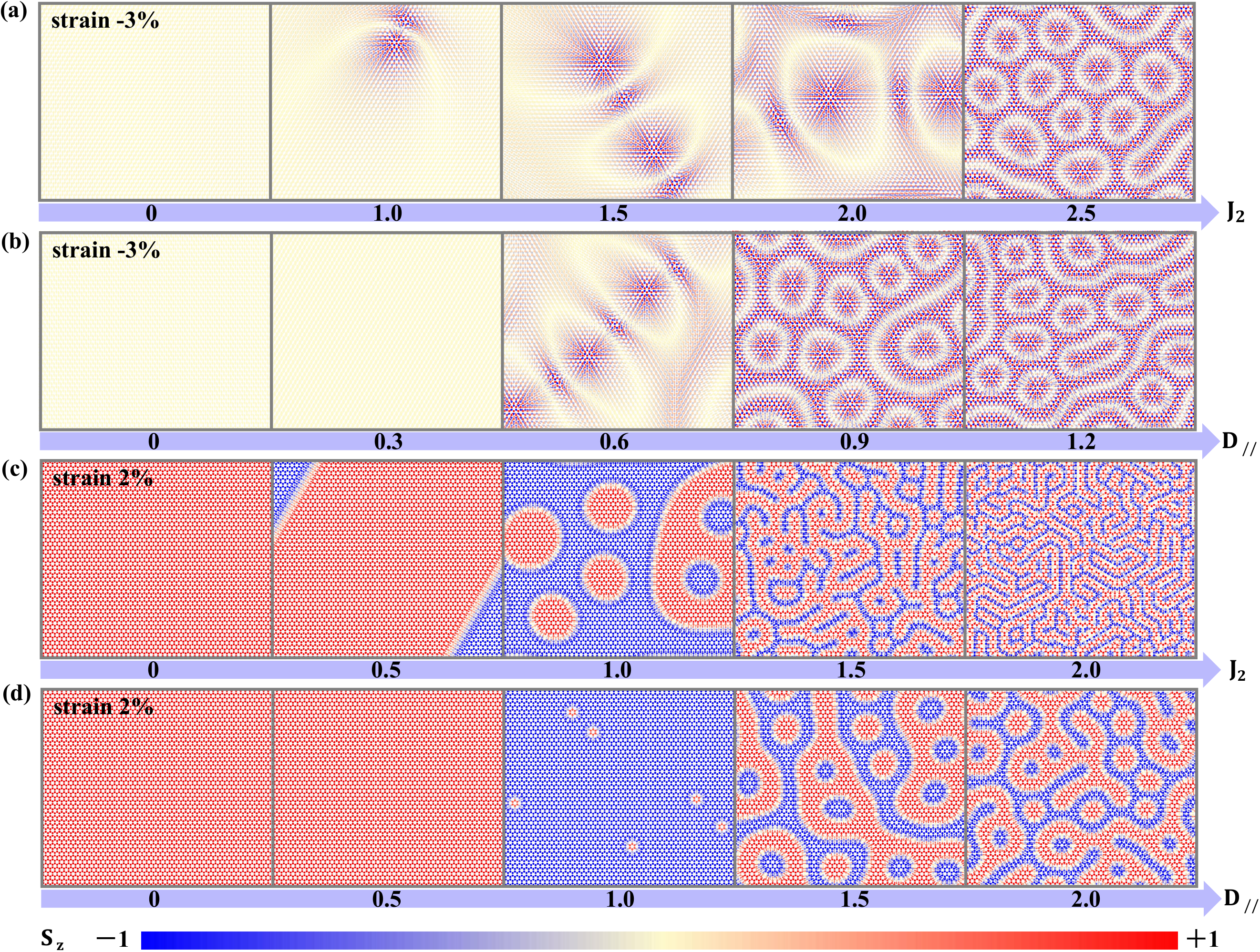}
\caption{\label{fig:figure4} 
Magnetic configurations as functions of the $J_2$ and $D_{//}$ under different strain conditions.
(a) Evolution of spin textures with varying 2NN exchange coupling $J_2$ for the $-3\%$ strained structure.
(b) Evolution of spin textures with increasing in-plane DMI component $D_{//}$ for the $-3\%$ strained structure.
(c) Magnetic configurations as a function of $J_2$ for the $+2\%$ strained structure.
(d) Magnetic configurations as a function of $D_{//}$ for the $+2\%$ strained structure.
All energy parameters are given in units of meV/Cr. The color scale represents the out-of-plane spin component, where red and blue correspond to spin-up and spin-down orientations, respectively. Only one-quarter of the full simulation area is displayed for clarity.
}
\end{figure}

To elucidate the microscopic mechanism governing the stabilization of AFM and FM skyrmions, we performed a comparative energy analysis based on the spin Hamiltonian introduced above. Specifically, we evaluated the energy difference between the collinear N$\mathrm{\acute{e}}$el-type AFM (FM) ground state and the corresponding AFM (FM) skyrmionic state, and decomposed these energy differences into distinct magnetic contributions, including the Heisenberg exchanges ($J_i$), SIA energy ($A_{zz}$), and DMI ($D_z$, $D_{//}$). The results, summarized in Table~\ref{tab:table1}, reveal that the total energy differences are small for both AFM and FM skyrmions, suggesting that these topological states are energetically accessible in the same 2D Janus system.
For the $-3\%$ strained system hosting AFM skyrmions, as shown in Table~\ref{tab:table1}, both $J_2$ and $D_{//}$ yield negative values of $\Delta E_1$, indicating that they contribute to overall energy gain and play crucial roles in stabilizing the AFM skyrmion state. To elucidate how these parameters govern the resulting spin configurations, we performed a series of LLG simulations by varying the magnetic interactions. First, $J_2$ is tuned from 0 to 2.5 meV/Cr while keeping all other parameters fixed, and the magnetic configurations are illustrated in Fig.~\ref{fig:figure4}(a). When $J_2=0$, the spins lie predominantly in-plane without forming any topological texture. As $J_2$ increases to approximately 1 meV/Cr, AFM bimerons begin to appear, and well-defined AFM skyrmions emerge near $J_2 = 2.5$ meV/Cr. Next, to examine how $D_{//}$ affects the magnetic configuration, we vary its value from 0 to 1.2 meV/Cr while fixing other parameters, as shown in Fig.~\ref{fig:figure4}(b). At $D_{//} = 0$, the spins remain nearly collinear within the plane. When $D_{//}$ reaches about 0.6 meV/Cr, AFM bimerons appear, consisting of two oppositely polarized merons in the two AFM-coupled sublattices. Upon further increasing $D_{//}$ to 0.9 meV/Cr, fully developed AFM skyrmions become stabilized. These results confirm that both moderate $J_2$ and sizable in-plane DMI cooperatively promote the formation of AFM skyrmions by introducing exchange frustration and chiral twisting.

As for the FM skyrmion state, Table~\ref{tab:table1} shows that both $J_2$ and $D_{//}$ also produce negative values of $\Delta E_2$, suggesting that they likewise contribute to the energy stabilization of FM skyrmions. To further verify their effects, we performed similar parameter-tuning LLG simulations. As shown in Fig.~\ref{fig:figure4}(c), $J_2$ is varied from 0 to 2 meV/Cr while keeping all other parameters fixed. At $J_2 = 0$, the system remains uniformly ferromagnetic. A slight increase in $J_2$ induces stripe-like modulations, and stable FM skyrmions appear near $J_2 = 1$ meV/Cr. Their diameter decreases gradually as $J_2$ increases further, indicating that moderate AFM $J_2$ enhances exchange frustration and promotes smaller skyrmion formation. Similarly, when $D_{//}$ is varied from 0 to 2 meV/Cr, as shown in Fig.~\ref{fig:figure4}(d), the magnetic structure tends to become noncollinear. FM skyrmions emerge once $D_{//}$ exceeds 1 meV/Cr, and their diameter decreases progressively with increasing DMI strength, consistent with enhanced chiral twisting of spins. These results demonstrate that the cooperative effects of $J_1$, $J_2$, and $D_{//}$ are also essential for stabilizing FM skyrmions in the tensile-strained system.

\subsection{Magnetic textures under external magnetic field}

\begin{figure}
\centering
\includegraphics[width=0.9\textwidth]{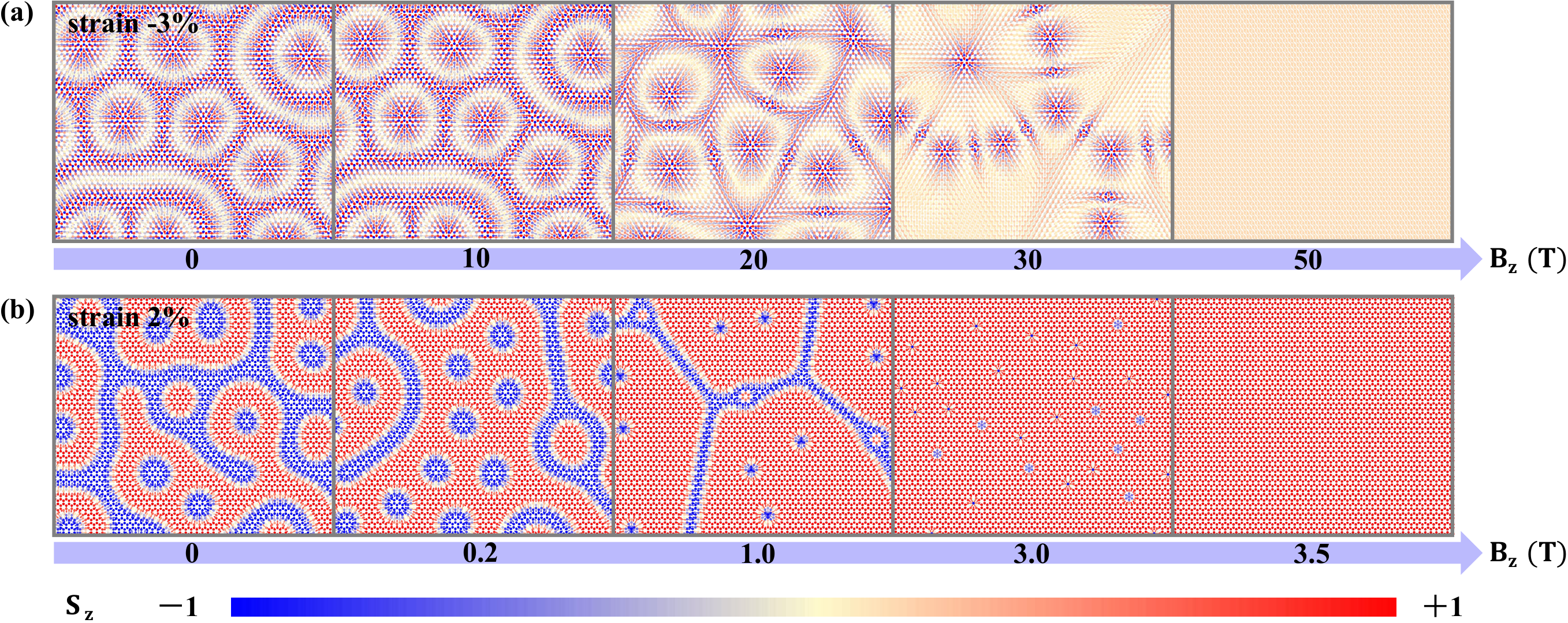}
\caption{\label{fig:figure5} 
Evolution of magnetic textures under an external magnetic field ($B_z$) in monolayer $\ce{Cr2Ge2Te3S3}$.
(a) Field-dependent evolution of the AFM skyrmions phase at $-3\%$ strain under increasing out-of-plane magnetic field.
(b) Field-dependent evolution of the FM skyrmions phase at $+2\%$ strain under increasing out-of-plane magnetic field.
Only one-quarter of the full simulation area is displayed for clarity.
}
\end{figure}

\begin{table}
\caption{\label{tab:table1} Upper: Total energy of AFM skyrmion (AFMsk) state and N$\mathrm{\acute{e}}$el-type AFM state for $\ce{Cr2Ge2Te3S3}$ under $-3\%$ strain, as well as the decomposed energies into each term of Hamiltonian. Energy unit is meV/u.c.. The ground state corresponds to the AFMsk configuration, as shown in Fig.~\ref{fig:figure3}(c1).
Lower: Total energy of FM skyrmion (FMsk) state and FM state, as well as the decomposed energies into each term of Hamiltonian. Energy unit is meV/u.c.. The ground state corresponds to the FMsk configuration, as shown in Fig.~\ref{fig:figure3}(d1).
The third row of upper and lower sections reports the energy difference, which are $\Delta E_1=E_{\text{AFMsk}}-E_{\text{AFM}}$ and $\Delta E_2=E_{\text{FMsk}}-E_{\text{FM}}$. The above calculations are performed in a $100 \times 100 \times 1$ supercell containing $2\times 10^4$ Cr atoms.
}
\begin{tabular}{cccccccc}
\hline\hline
 & Total & $J_1$ & $J_2$ & $J_3$ & $D_{//}$ & $D_z$ & $A_{zz}$ \\
\hline
AFMsk state & -0.925 & -1.295 & 0.678 & -0.294 & -0.017 & -0.002 & 0.006 \\
AFM state & -0.908 & -1.325 & 0.723 & -0.321 & 0 & 0 & 0.015 \\
$\Delta E_1$ & -0.018 & 0.029 & -0.045 & 0.027 & -0.017 & -0.002 & -0.009 \\
\hline\hline
FMsk state & -0.684 & -0.763 & 0.297 & -0.045 & -0.045 & 0 & -0.034 \\
FM state & -0.681 & -0.804 & 0.348 & -0.173 & 0 & 0 & -0.052 \\
$\Delta E_2$ & -0.003 & 0.041 & -0.051 & 0.033 & -0.045 & 0 & 0.018\\
\hline\hline
\end{tabular}

\end{table}

The application of an external magnetic field provides an effective route for manipulating topological spin textures \cite{PhysRevB.102.241107, PhysRevB.101.184401, PhysRevB.98.060413}. In chiral magnets dominated by FM exchange couplings, a perpendicular magnetic field is known to stabilize skyrmions through the delicate competition among exchange interactions, magnetic anisotropy, and DMI \cite{doi:10.1021/acsnano.2c08544, PhysRevB.92.214439}. In contrast, the magnetic-field response of AFM skyrmions has received far less attention. Understanding how AFM skyrmions evolve under an out-of-plane magnetic field ($B_z$) and their comparison with FM skyrmions is crucial for the design of field-tunable topological spintronic devices.
To explore this, we performed Landau-Lifshitz-Gilbert (LLG) micromagnetic simulations for the $-3\%$ strained monolayer, where AFM skyrmions are stabilized in the absence of an external field. When a weak perpendicular field ($B_z = 10$ T) is applied, as shown in Fig.~\ref{fig:figure5}(a), the spin configuration remains nearly unchanged, indicating the remarkable robustness of AFM skyrmions against magnetic perturbations. As the magnetic field increases to $B_z = 20$ T, the AFM skyrmions gradually transform into AFM merons. This field-induced skyrmion-to-meron transition differs fundamentally from that in FM systems, in which increasing the magnetic field usually drives the merger of meron pairs into complete skyrmions \cite{https://doi.org/10.1002/adfm.202104452}. Upon further field increase, the AFM merons shrink, their density decreases, and eventually the system transitions into a nearly coplanar spin configuration as the Zeeman energy overtakes the DMI and exchange terms.

We also investigated the magnetic-field evolution of FM skyrmions stabilized under $+2\%$ strain. At moderate fields ($B_z = 3$ T), as shown in Fig.~\ref{fig:figure5}(b), the system exhibits a dense array of small-sized FM skyrmions. As the field increases to $B_z = 3.5$ T, the skyrmionic lattice collapses into a uniform FM state. Given that AFM skyrmions only start to show quantitative changes under a $20$ T magnetic field, this behavior demonstrates that FM skyrmions are considerably less resistant to magnetic fields than their AFM counterparts. The contrasting field responses also highlight the intrinsic robustness of AFM skyrmions, which maintain topological stability even under strong magnetic perturbations. Taken together with the strain-induced switching discussed above, these results establish that both AFM and FM skyrmions can be controlled within a single Janus monolayer through the combined application of strain and magnetic fields. 

\section{Conclusion}
In summary, our first-principles calculations and atomistic spin simulations demonstrate that it is possible to switch between AFM and FM skyrmion phases within the same system through strain engineering, as exemplified by the Janus monolayer $\ce{Cr2Ge2Te3S3}$. The two topological states arise from the delicate competition among exchange interactions and DMI, and can be interconverted by tuning the magnetic coupling strength. Strain provides an effective means of tuning the magnetic coupling strength, where compressive strain stabilizes AFM skyrmions and tensile strain favors FM skyrmions. Both phases exhibit dynamical and thermal stability and respond differently to magnetic fields, with FM skyrmions being more easily modulated while AFM skyrmions remain robust.
These results establish a general framework for achieving and controlling different topological magnetic states within a single 2D magnet. Beyond demonstrating the strain-induced switching between AFM and FM skyrmions, this work reveals the connection between exchange frustration and chiral interactions, offering a versatile platform for reconfigurable topological spintronic devices.

\section*{Acknowledgements}
The authors thank the National Key R\&D Program of China (2023YFA1406400) for financial support.


\providecommand{\latin}[1]{#1}
\makeatletter
\providecommand{\doi}
  {\begingroup\let\do\@makeother\dospecials
  \catcode`\{=1 \catcode`\}=2 \doi@aux}
\providecommand{\doi@aux}[1]{\endgroup\texttt{#1}}
\makeatother
\providecommand*\mcitethebibliography{\thebibliography}
\csname @ifundefined\endcsname{endmcitethebibliography}  {\let\endmcitethebibliography\endthebibliography}{}

\end{document}